\documentclass[twocolumn,showpacs,preprintnumbers,amsmath,amssymb,prl]{revtex4}

\usepackage{graphicx} \usepackage{dcolumn} \usepackage{bm}
\usepackage{textcomp}

\newcommand{\Em}{\ensuremath{E_m}}
\newcommand{\Ecx}{\ensuremath{E_{cx}}}
\newcommand{\Ek}{\ensuremath{E_\mathrm{co}}}
\newcommand{\Ech}{\ensuremath{E_\mathrm{ch}}}
\newcommand{\nl}{\ensuremath{n_l}}
\newcommand{\nr}{\ensuremath{n_r}}
\newcommand{\nx}{\ensuremath{n_x}}
\newcommand{\ny}{\ensuremath{n_y}}
\newcommand{\ns}{\ensuremath{n_s}}
\newcommand{\dn}{\ensuremath{\Delta{}n}}
\newcommand{\Gmax}{\ensuremath{G_{\mathrm{max}}}}
\newcommand{\nmax}{\ensuremath{n_{\mathrm{max}}}}
\newcommand{\Gmin}{\ensuremath{G_{\mathrm{min}}}}
\newcommand{\Gm}{\ensuremath{G_{m}}}
\newcommand{\DGqf}{\ensuremath{\Delta G_{\mathrm{qf}}}}
\newcommand{\Csl}{\ensuremath{C_{sl}}}
\newcommand{\Csr}{\ensuremath{C_{sr}}}

\begin{document}

\title{Coulomb blockade in two island systems with highly conductive junctions}

\author{Bernhard Limbach}
\author{Peter vom Stein}
\author{Christoph Wallisser}
\author{Roland Sch{\"a}fer}
\affiliation{
Forschungszentrum Karlsruhe, Institut f{\"u}r Festk{\"o}rperphysik\\ Postfach
3640, 76021 Karlsruhe, Germany }

\date{\today}

\begin{abstract}
We report measurements on single-electron pumps, consisting of two metallic
islands formed by three tunnel junctions in series.  We focus on the
linear-response conductance as a function of gate voltage and temperature of
three samples with varying system parameters.  In all cases, strong quantum
fluctuation phenomena are observed by a $\log\left(k_BT/(2\Ek)\right)$
reduction of the maximal conductance, where $\Ek$ measures the coupling
strength between the islands. The samples display a rich phenomenology,
culminating in a non-monotonic behavior of the maximal conductance as a
function of temperature.
\end{abstract}



\maketitle

The transport properties of single electron devices are well described by the
so called sequential tunneling model as long as the conductances of the
underlying tunneling contacts are small compared to the conductance quantum
$G_K=e^2/h$ \cite{averin91,ingold92}.  In most practical cases, however, one
has to include higher-order corrections for a quantitative description due to
quantum fluctuations of the charge number on the single electron islands. This
has been demonstrated clearly for the single electron transistor (SET)
\cite{koenig97,joyez97,koenig98a,chouvaev99,christophpaper}.  Excellent
agreement between experimental investigations and theoretical studies has been
established at low conductances with the aid of perturbation expansion
including the correct description of a logarithmic contribution to the
linear-response conductance of the multichannel-Kondo type.  At higher
conductances however, perturbation theory breaks down.  Nevertheless, by
applying Monte-Carlo methods, the experimental results can be described with
amazingly high accuracy \cite{joyez97,goeppert00,christophpaper}.  The findings
on the SET indicate that a close match between the model Hamiltonian and its
experimental realization exists for single electron devices.  It is not clear a
priori whether the good agreement between perturbation expansion and experiment
as well as Quantum-Monte-Carlo numerics found for the SET survives if more
complex arrangements of single electron islands are investigated.  E.\ g.\ the
low temperature conductance of the SET involves finite occupation of only two
states and the mapping onto the multichannel-Kondo model is based on
identifying those two states with a pseudo spin. This procedure has no simple
analogy in general arrangements.  It is worthwhile to check the range of
validity of perturbation theory in more general cases.  Two island systems,
readily accessible both by experiment and theory, serve as a good starting
point. 
\begin{figure}
\includegraphics[width=0.95\linewidth]{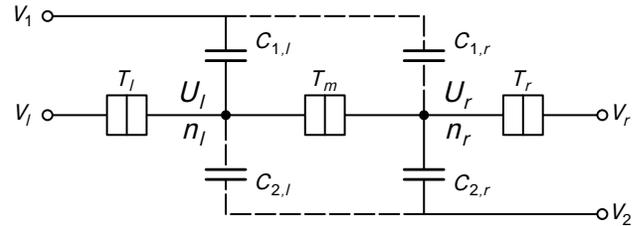}
\caption{\label{fig:1}
Schematic representation of the single-electron pump.  The three tunneling
contacts $T_l$, $T_m$, and $T_r$ are laid out in a row forming two islands
(each contact $i\in\{l,m,r\}$ is characterized by its capacitance $C_i$ and
conductance $G_i\equiv{}g_iG_K$). \nl{} and \nr{} count the number of electrons
by which the left and right island charge differs from neutrality.  The
arrangement is biased by the voltage difference $V_r-V_l$.  The electrostatic
potentials on the islands ($U_l$ and $U_r$) can be tuned by $V_1$ and $V_2$
which couple directly to the islands by the capacitances $C_{1l}$ and $C_{2r}$.
$C_{2l}$ and $C_{1r}$ represent the experimentally unavoidable stray
capacitances.
}
\end{figure}
\begin{figure}
\includegraphics[width=\linewidth]{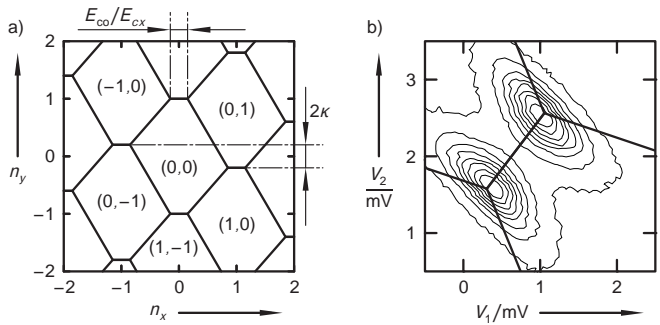}
\caption{\label{fig:2}
a) Stability diagram for the SEP sketched in Fig.\ \ref{fig:1}.  The hexagonal
cells mark regions where the indicated charge state $(\nl,\nr)$ possesses the
lowest energy $\Ech(\nl,\nr)$. b) Measurement of the linear-response
conductance of sample 2 as a function of both gate voltages $V_1$ and $V_2$ at
$150\,$mK.  The outermost contour line indicates a conductance of
$0.01\,$\textmu{}S.  The following lines range from $0.05\,$\textmu{}S to
$0.4\,$\textmu{}S with a $0.05\,$\textmu{}S spacing.  The stability diagram
(thick lines) is deformed in the coordinates of this figure.
}
\end{figure}

In this Letter we present an experimental study of the linear-response
conductance of two single electron islands in series, an arrangement nicknamed
single electron pump (SEP) \cite{pothier91,pothier92,lotkhov01} and sketched in
Fig.\ \ref{fig:1}.  The linear-response conductance varies with the gate
voltages and is bound between temperature dependent values $\Gmin(T)$ and
$\Gmax(T)$.  For samples in an interesting and  accessible parameter range, the
maximal conductance \Gmax{} obeys a non-monotonic behavior as a function of
temperature, which is  reported here for the first time.  This behavior,
astonishing at first sight, is naturally explained by the sequential tunneling
model in this Letter, giving evidence that non trivial behavior is to be
expected in complex arrangements of single-electron islands. Beyond the
validity range of the sequential model,  we find a logarithmic correction of
the conductance at low temperature due to quantum fluctuations. Our high
quality data make a sensitive test of perturbation theory as developed e.\ g.\
in Ref.\ \onlinecite{pohjola99} feasible. This however, requires an elaborate
calculation along the line of formulae in Ref. \onlinecite{pohjola99}, which is
out of the scope of the work presented here.

The linear-response conductance of the SEP can be modeled in various coordinate
systems spanning the $(V_1,V_2)$ plane.  In this letter we choose dimensionless
coordinates \nx{} and \ny{} defined such that the charging energy
$\Ech=\Ecx(\nx-n_s)^2 +\Em(\ny+\dn+\kappa\ns)^2$. Here $\Ecx=e^2/(2C_s)$ with
$C_s$ being the sum of all external capacitances in Fig.\ \ref{fig:1} (i.e. all
capacitances except $C_m$), $\Em=(e^2/2)C_s/(C_mC_s+\Csl\Csr)$ with \Csl{} and
\Csr{} denoting the sum of the external capacitances on the left and the right
island separately, and $\kappa=(\Csr-\Csl)/C_s$. The coordinate \nx{} is
associated with the change of the total charge number $\ns=\nl+\nr$, while
\ny{} redistributes the charge between both islands ($\dn=\nl-\nr$).

Each point in the plane spanned by \nx{} and \ny{} can be mapped onto a charge
ground state $(\nl,\nr)$ which gives the lowest possible electrostatic charging
energy $\Ech(\nl,\nr)$.  This procedure divides the $(\nx,\ny)$ plane into the
grid of hexagonal cells depicted in Fig.\ \ref{fig:2}a.  The length $\Ek/\Ecx$
of the horizontal cell boundaries is a measure of the coupling strength, where
$\Ek=\Ecx+\Em(\kappa^2-1)/4\propto{}C_m$. Due to the periodicity it suffices to
study the linear-response conductance in a small exemplary portion of the
$(\nx,\ny)$ plane.  Within the sequential model, the linear-response
conductance of the system vanishes exponentially at low temperatures except
close to the triple points in the $(\nx,\ny)$ plane, where the ground-state
energy of all three adjacent states is degenerate.  The conductance peaks near
these points.  It is worthwhile to mention a peculiar behavior of the SEP:
Exactly at the triple points the low temperature conductance is constant and
given by $G_0/3$ where $G_0^{-1}=G_l^{-1}+G_r^{-1}+G_m^{-1}$.  Although for
$T\ll\Ek/k_B$ the maximal conductance \Gmax{} is temperature independent as
well, the system assumes \Gmax{} at a slightly different position in the \nx{}
direction \cite{pohjola99}.  Taking only three states into account (the
occupation probability of all other states is exponentially small for
$T<\Ek/k_B$) we get
\begin{equation}\label{eq:thpe}
G(\nx)|_{\ny=-\kappa}=\frac{G_s}{2+e^{-\beta\Delta{}E}}
\left(\frac{g_s}{g_m}+\frac{e^{\beta\Delta{}E}-1}{\beta\Delta{}E}\right)^{-1},
\end{equation}
with $\beta=1/(k_BT)$, $g_s=g_lg_r/(g_l+g_r)=G_s/G_K$, and
$\Delta{}E=2\Ecx(\nx-1)+\Ek$.  Depending on the ratio $g_s/g_m$ the peak
position deviates from the triple point at $\Delta{}E=0$ for $g_m\not=2g_s$. 

In a measurement as a function of \nx{} and \ny{} the conductance displays a
periodic grid of peaks (grouped as pairs), marking the endpoints of the
horizontal boundaries in Fig.\ \ref{fig:2}a.  With rising temperature the peaks
broaden and shift towards the center of the horizontal boundaries.  Fig.\
\ref{fig:2}b gives an example  at $T=150\,$mK$\sim0.09\Ecx/k_B$, where thermal
broadening already is effective in merging the two separate peaks.  Finally the
two peaks merge completely and the conductance takes its maximal value at the
mid-points, e.\,g.\ at $\nx=1$ and $\ny=-\kappa$.

\begin{figure}
\includegraphics[width=2.1566666666666667in]{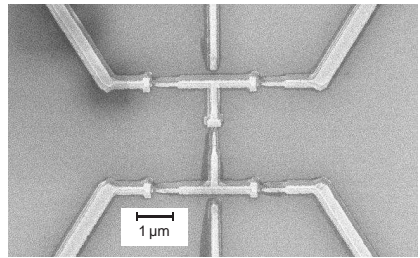}
\caption{\label{fig:3}
Scanning electron microscope picture of sample 2.  The two T-shaped structures
in the middle are the islands.  The inner tunneling contact is visible in the
center of the picture.  The outer contacts are turned by $90^\circ$ with
respect to the former.  The gate lines are visible at top and bottom pointing
up and down towards the inner contact.  In the final measurement the top and
bottom leads are biased in parallel.
}
\end{figure}

We study three samples in two different layouts.  All samples have been
produced by standard shadow-evaporation technique from aluminum with
aluminum-oxide barriers.  The barriers for the middle and outer contacts are
fabricated in different oxidation steps,  making different barrier thicknesses
for internal and external contacts possible.  Sample 1 has a simple layout (not
shown) with three contacts in a row, which is the most natural arrangement to
fabricate a SEP. In this layout the serial conductance is accessible, but the
conductance of the individual contacts remains unknown. In the slightly more
complex structure of sample 2 and 3 (see Fig.\ \ref{fig:3}) each island is
connected to independent leads via two contacts. This permits measuring the
conductance of different serial combinations of the contacts, and thus the
individual contact conductances can be determined.  In the final experiments
both external contacts of each island are operated in parallel (connected to
the same voltage source), and the two contacts then act exactly as a single
contact.  We could not detect any degradation (considering noise performance or
sensitivity to external disturbances) for the latter samples compared to sample
1.

From the positions of the conductance peaks in the $(\nx,\ny)$ plane at low
temperatures (see Fig. \ref{fig:2}) we obtain the parameters $\kappa$ and
$\Ek/\Ecx$.  In addition we measure the conductance in the high temperature
region where it does not depend on either \nx{} nor on \ny.  In close analogy
to the high temperature expansion for the SET, it behaves as
$G(T)\approx{}G_0\left((1-E_c/(3k_BT)\right)$ .  It can be shown
\cite{hirvi95,limbach02} that the relation 
$$\Ecx=\frac{E_c}{G_0}\left(\frac{e_+}{G_l}+\frac{e_-}{G_r}+
\frac{e_++e_--2\Ek/\Ecx}{G_m}\right)^{-1}$$
holds, where 
$e_\pm=\left(\left(\Ek/\Ecx\right)\left(\kappa\pm1\right)
\mp2\right)/\left(\kappa\mp1\right)$
and $E_c$ is an experimentally determined
fitting parameter.  Tab.\ \ref{tb:1} gives all relevant sample parameters.  

\begin{table}
\caption{\label{tb:1}Parameters of the three samples. }
\begin{tabular}{llllllllll}
\hline
samp.& $\displaystyle\frac{\Ecx}{k_B}$&$\displaystyle\frac{\Em}{k_B}$
&$\displaystyle\frac{\Ek}{k_B}$&$\displaystyle\frac{E_c}{k_B}$& $\kappa$&
$g_l$&	$g_r$&	$g_m$&	$G_0$\\
 &	(K)&	(K)&	(K)& (K)&&&&&(\textmu{}S)\\
\hline\hline
1& 2.8&5.8&1.3&5.6&-0.018&0.44&0.44&0.04&1.42\\
2& 1.6&3.0&0.9&2.5&0.10&0.52&0.83&1.32&10.0\\
3& 1.5&4.7&0.3&4.5&0.0013&0.73&0.57&0.03&0.95\\
\hline
\end{tabular}
\end{table}

\begin{figure}
\includegraphics[width=0.95\linewidth]{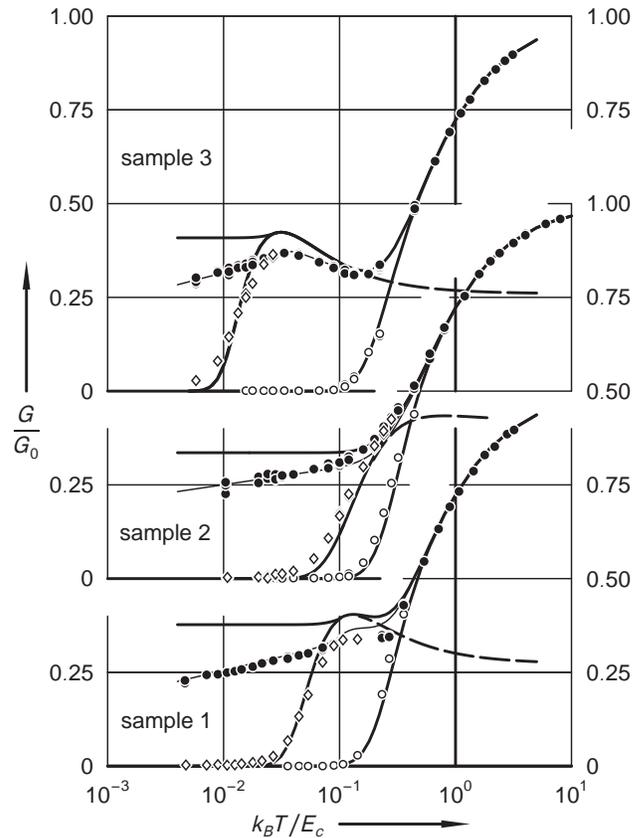}
\caption{\label{fig:4}
Linear response conductance of the three samples as a function of temperature.
Shown are \Gmin{} ({\normalsize $\circ$}), \Gmax{} ({\normalsize $\bullet$}),
and \Gm{} ({\scriptsize $\lozenge$}, the conductance at $n_x=0.5$). The dashed
lines are the result of an analytical solution of a 4 state model (see text).
As solid lines the outcome of the sequential model with parameters from Tab.\
\ref{tb:1} is shown.  For the thin solid line a term of the form
$\alpha\log(k_BT/\Ek)$ has been added to the maximal conductance as calculated
from the sequential model.  Here $\alpha$ is a fitting parameter.  
}
\end{figure}

To simplify the analysis of our data and facilitate the comparison with
theoretical considerations we focus on the temperature dependence of three
special values of the conductance in the $(\nx,\ny)$ plane, namely \Gmin,
\Gmax, and \Gm, the latter defined as the conductance at the center of the
horizontal boundaries in Fig.\ \ref{fig:2}a (i.\,e. at $\nx=1,\ny=-\kappa$).
In Fig.\ \ref{fig:4} we display our main findings.  In addition to our
experimental results we present the outcome of a calculation in the framework
of the sequential tunneling model using the parameters from Tab.\ \ref{tb:1}.
The techniques for such a calculation are well documented in the literature
(e.\,g.\ \cite{ingold92}) so we do not comment on this calculation in detail.
It requires the solution of a master equation which one may set up using golden
rule rates for the inelastic tunneling events.  For $T\ll E_c/k_B$ at most four
$(\nl, \nr)$ states are occupied with reasonable probability.  Restricting the
master equation to these four states allows for an analytical solution.  At
higher temperatures numerical relaxation methods are used.

The sequential tunneling model gives good agreement with our experimental data
with the exception of \Gmax{} at low temperature. The latter deviations are
discussed at the end of the letter.  The overall behavior (see Fig.
\ref{fig:4}) is governed by two scales ($\Ek/k_B$ and $E_c/k_B$) at which
$\Gm(T)$ and $\Gmin(T)$, respectively, start to increase and finally merge with
$\Gmax(T)$.  To get finite conductance through the SEP at least three charge
states have to be occupied in thermal equilibrium.  At $T\ll\Ek/k_B$ this is
only possible near the triple points where three adjacent states are occupied.
Here e.\,g.\ the sequence $(0,0)\to(1,0)\to(0,1)\to(0,0)$, corresponding to a
charge transfer from left to right, occurs with finite probability.  The
inverse process is equiprobable, but under voltage bias a net current occurs.
At $\nx=1,\ny=-\kappa$ where \Gm{} is measured, the charging energy of states
$(0,0)$ and $(1,1)$ lie $\Delta{}E=\Ek$ above the two fold degenerate ground
state ($(0,1),(1,0)$).  As a result the linear response conductance at
$T\ll\Ek/k_B$, as calculated from the sequential model, is exponentially
suppressed since no charge transfer is possible using only two states.  At
$T\sim\Ek/k_B$ the states $(0,0)$ and $(1,1)$ are thermally occupied with
increasing probability leading first to an exponential increase of \Gm{} and
finally to the merging of \Gm{} and \Gmax{}.  For $\nx=\ny=0$ states besides
$(0,0)$ are occupied for $T\gtrsim{}E_c/k_B$ only.  Thus \Gmin{} is
exponentially small for $T\ll{}E_c/k_B$.

The most striking feature of our measurements is the non-monotonic dependence
of \Gmax{} on the temperature found for sample 3.  It is clear from Fig.\
\ref{fig:4}  that the phenomenology is correctly described by the sequential
model.  To get more insight into the nature of the drop of \Gmax{} at
$T>\Ek/k_B$ we analyze $\Gm(T)$, which coincides with \Gmax{} in the relevant
temperature range.  The four-state approximation mentioned above yields an
analytical solution:
\begin{equation}\label{eq:gmfs}
\Gm^{(4)}=\frac{G_s}{2}\frac{\beta\Ek}{\sinh(\beta\Ek)}
\left(1-\frac{2g_s}{g_m}\frac{\beta\Ek}{1-e^{\beta\Ek}}\right)^{-1}.
\end{equation}
We have drawn this function for our sample parameters in Fig.\ \ref{fig:4} as
dashed lines. It has a distinct maximum for $g_m<g_s$, the position and
strength of which depends on the ratio $g_s/g_m$. The above approximation
breaks down at $T\sim{}E_c/k_B$ where more than four states are occupied
significantly resulting in a rapid rise of the conductance.  For an
experimental observation of a local minimum the relation $\Ek<E_c$ has to be
fulfilled in addition to $g_m<g_s$.

\begin{figure}
\includegraphics[width=0.65\linewidth]{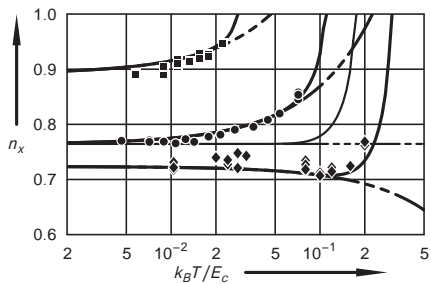}
\caption{\label{fig:5}
Position of the conductance maxima as a function of temperature.
{\normalsize$\bullet$}: sample 1, {\scriptsize$\blacklozenge$}: sample 2,
{\tiny$\blacksquare$}: sample 3.  Lines: Predictions of the full sequential
model (solid) and Eq.\ \ref{eq:thpe} (three-state approximation, dash dotted).
For sample 1 the peak position has been used to adjust $g_s/g_m=5$.  Assuming
$g_l=g_r=g_m$ gives the thin lines failing to fit the data.
}
\end{figure}

In Fig. \ref{fig:5} we plot the \nx{} coordinate of the conductance maximum
\nmax{} as a function of temperature.  Again, reasonable agreement with the
sequential model is found.  For $T\to0$ the position of the maximum approaches
the location of the triple point.  At low temperature the conductance peak is
described by Eq.\ \ref{eq:thpe}.  The maximum can shift in either direction
depending on the ratio $g_s/g_m$.  This can be used to determine $g_s/g_m$ for
sample 1 ($g_s/g_m=5$) where this ratio can not be measured directly due to the
simple layout (see above).

The sequential model fails to predict the temperature dependence of $\Gmax$ at
low temperature.  This deviation can be described phenomenologically by adding
a term of the form $\DGqf=\alpha\log(k_BT/(2\Ek))$.  The thin solid lines in
Fig.\ \ref{fig:4} display the outcome of a fitting procedure in $\alpha$ that
minimizes the mean square deviation between the measured values of \Gmax{} and
$G_{\mathrm{seq}}+\DGqf(\alpha)$, showing very good agreement.  We find
$\alpha_1=45\,$nS, $\alpha_2=200\,$nS, and $\alpha_3=33\,$nS for sample 1, 2,
and 3, respectively.  For the SET much the same behavior is observed and
attributed to quantum fluctuations of the charge states.  Pohjola et al.\
\cite{pohjola99} analyzed the linear response of the SEP by renormalization
group methods. They also found a logarithmic behavior of the low temperature
conductance, in qualitative agreement with our experimental result.  However,
quantitative results were obtained for certain limiting cases of special
interest only.  For a detailed comparison with our experiment a calculation
using our sample parameters is highly desirable.

In summary we have presented an experimental study of the linear response
conductance of the SEP in a regime where quantum fluctuations of the charge
eigenstates can not be ignored.  Depending on the ratio $g_s/g_m$ the SEP shows
a remarkably rich phenomenology even within the framework of lowest-order
perturbation theory, the so-called sequential tunneling model. Most strikingly,
in the easily accessible regime $\Ek<\Ecx$ and $g_m\ll2g_s$ a pronounced
non-monotonic temperature dependence of the conductance has been observed.
Phenomena of this kind are to be expected in all single electron devices which
are more complex than the SET. They can uncover internal characteristics
unaccessible by other means -- here   e.\ g.\ $g_s/g_m$ which is not directly
measurable in the most natural SEP layout (sample 1).  At low temperature
deviations from the sequential behavior due to quantum fluctuations become
clearly visible.  They are described in close analogy to the SET by a
logarithmic correction term of the form $\DGqf=\alpha\log(k_BT/(2\Ek))$, in
qualitative agreement with the findings of Ref. \onlinecite{pohjola99}.  We
propose a reevaluation of the formulae of Ref.\ \onlinecite{pohjola99} with
parameters in accordance with our experiment so as to check  the applicability
of perturbation expansion for devices more complex than the SET.

We acknowledge useful discussions with G.\ G{\"o}ppert, G.\ Johannson, P.\
Joyez, J.\ K{\"o}nig, H.\ Pothier, and H.\ v.\ L{\"o}hneysen. This work has
been carried out as part of SFB 195.

\bibliography{cbitis}

\end{document}